\documentclass[preprint]{aastex631}

\usepackage{natbib}

\begin{document}

\title{A Predicted Great Dimming of T~Tauri: Has it Begun?}

\author[0000-0002-6881-0574]{Tracy L. Beck}
\affiliation{Space Telescope Science Institute\\
3700 San Martin Drive \\
Baltimore, MD 21218, USA}

%% Note that the \and command from previous versions of AASTeX is now
%% depreciated in this version as it is no longer necessary. AASTeX 
%% automatically takes care of all commas and "and"s between authors names.

%% AASTeX 6.31 has the new \collaboration and \nocollaboration commands to
%% provide the collaboration status of a group of authors. These commands 
%% can be used either before or after the list of corresponding authors. The
%% argument for \collaboration is the collaboration identifier. Authors are
%% encouraged to surround collaboration identifiers with ()s. The 
%% \nocollaboration command takes no argument and exists to indicate that
%% the nearby authors are not part of surrounding collaborations.

%% Mark off the abstract in the ``abstract'' environment. 
\begin{abstract}
The optical star in the T~Tauri triple system is the prototype of young sun-like stars in our galaxy.  This complex and dynamic system has evidence for misaligned disks and outflows, and molecular material in a circumbinary ring that obscures the southern infrared binary, T~Tau~South.  Observations by members of the American Association of Variable Star Observers (AAVSO) show that T~Tau~North, the optical star, has dimmed by up to $\sim$2 magnitudes in the visual over the course of the past decade.  The dimming across the B, V, R and I bands has a color character typical of changes in ISM extinction, suggesting an increase in obscuration along the line of sight to T~Tau~North.  Material associated with the circumbinary ring around T~Tau~South has been predicted to occult the optical star via wide-scale orbital motion of the system. Through analysis of the geometrical configuration and motion of dust structures in the system, it seems that a great dimming of T~Tau~North by line-of-sight material associated with the T~Tau~South binary has, in fact, begun.  Based on the extent and motion of the circumbinary ring material associated with the southern binary, T~Tau~North will likely experience dimming events for decades to come and may disappear entirely from the optical sky as the densest mid-plane region of the ring traverses our line of sight.

\end{abstract}

%% Keywords should appear after the \end{abstract} command. 
%% The AAS Journals now uses Unified Astronomy Thesaurus concepts:
%% https://astrothesaurus.org
%% You will be asked to selected these concepts during the submission process
%% but this old "keyword" functionality is maintained in case authors want
%% to include these concepts in their preprints.
%\keywords{Classical T~Tauri Stars:}

%% From the front matter, we move on to the body of the paper.
%% Sections are demarcated by \section and \subsection, respectively.
%% Observe the use of the LaTeX \label
%% command after the \subsection to give a symbolic KEY to the
%% subsection for cross-referencing in a \ref command.
%% You can use LaTeX's \ref and \label commands to keep track of
%% cross-references to sections, equations, tables, and figures.
%% That way, if you change the order of any elements, LaTeX will
%% automatically renumber them.
%%
%% We recommend that authors also use the natbib \citep
%% and \citet commands to identify citations.  The citations are
%% tied to the reference list via symbolic KEYs. The KEY corresponds
%% to the KEY in the \bibitem in the reference list below. 

\section{Introduction} \label{sec:intro}
\begin{figure}[ht!]
\plotone{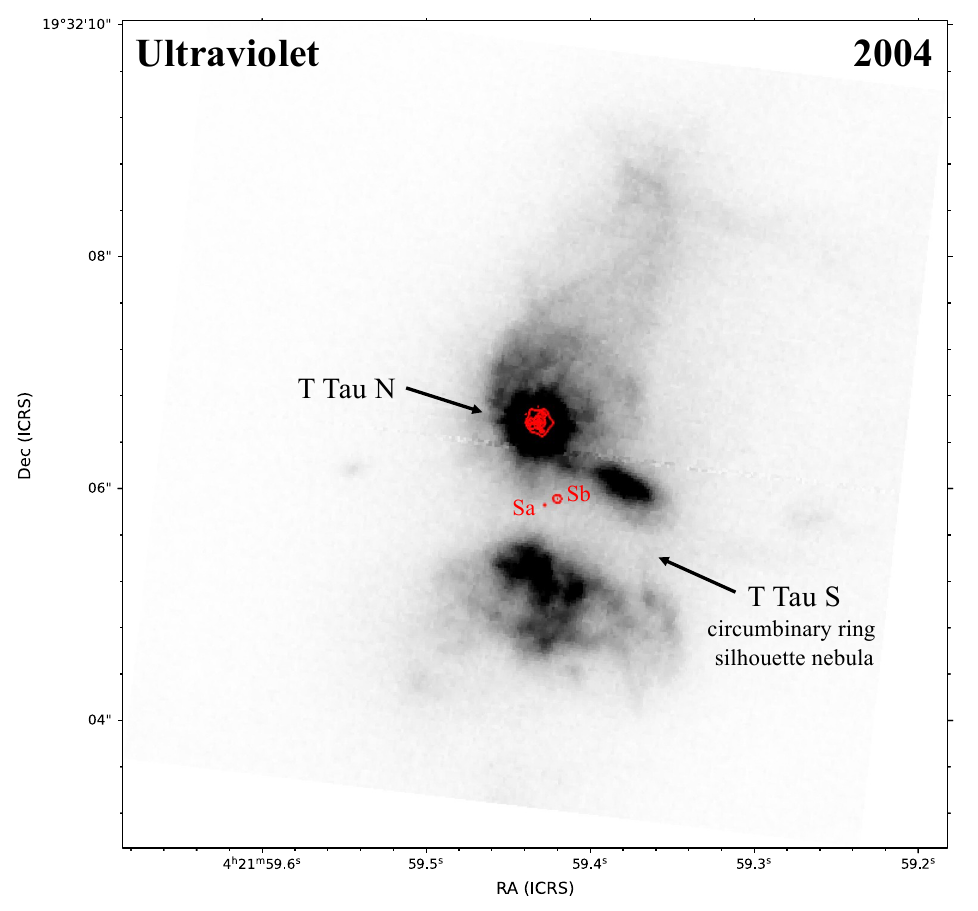}
\caption{The geometry of the T~Tauri triple system revealed in 2004 by UV imaging with the HST Advanced Camera for Surveys, with infrared contours in red showing the positions of the three stars (contour levels are 7, 15, 35, 70 and 90\% of the peak flux of T~Tau~N).  The image is log-scaled from 0 to 20\% of the peak flux to bring out detail in low level emission.  T~Tau~N is bright in UV flux, and the Sa and Sb binary stars are obscured to invisibility in the UV image by the circumbinary ring, which is viewed here in silhouette.  T~Tau~Sa was fainter than Sb in the infrared during this observation epoch in 2004.
\label{fig:fig1}}
\end{figure}

T~Tauri is a remarkable young triple star.  Historically, the optical star T~Tau~North (T~Tau~N) was highly variable in visible light prior to the 1920s, and then the variability suddenly stopped \citep{lozi49,beck01}.  The T~Tau~System has multiple Herbig-Haro outflows and a bright nebula, Hind's variable nebula, 30$"$ to the west \citep{hind64}.  T~Tau~South (T~Tau~S), the companion seen only at infrared and longer wavelengths, was discovered in 1982 \citep{dyck82}.   T~Tau~S was itself spatially resolved into two stars: Sa and Sb \citep{kore00}.   This $\sim$0.1$"$ average separation T~Tau~S binary is obscured to optical invisibility, even though it is located at a projected $\sim$0.$"$7 distance to the south of the bright optical star, T~Tau~N.   T~Tau~Sa+Sb is attenuated by $\sim$20 magnitudes of visual extinction \citep{gorh92, duch02, beck20}, from a circumbinary ring of foreground material oriented nearly edge-on (Figure~\ref{fig:fig1}; \cite{beck20}).  The circumstellar disk surrounding T~Tau~N is inclined by 28$^{\circ}$ and is not coplanar with the circumbinary ring, thus making this curious triple an ideal laboratory for studying dynamical evolution of mis-aligned disks in young multiple star systems \citep{beck20}.   

The motion of the Sa+Sb binary system has now been monitored through most of its orbit. Dynamical modeling and derivation of stellar masses shows that T~Tau~Sa is 2.05~$\pm$~0.14~M$_{\odot}$ and Sb is 0.43~$\pm$~0.06~M$_{\odot}$ \citep{kohl16,scha20}.  Based on evolutionary tracks and the optical spectral type K0, the mass of T~Tau~N is $\sim$2~M$_{\odot}$ \citep{scha20}. Hence, N and Sa are of comparable mass.  The optical and near-infrared flux of T~Tau~N was largely stable for the past several decades, yet T~Tau~Sa and Sb vary in the infrared by 3-4 magnitudes or more.  The character of this infrared variability is linked to the binary orbital motion, particularly for T~Tau~Sb \citep{beck20}.  The best-fit orbital period for Sa and Sb is $\sim$27 years and the binary experienced periastron closest approach in March 2023 \citep{scha20}.  The best-fit model orbit of T~Tau~N with respect to the center of mass of T~Tau~Sa~+~Sb has a period of 4600~years, though there is a range of acceptable model orbits for the wide system. With such a small arc of this wide orbit traced thus far, the possible orbital periods span a few hundred to over ten thousand years. \citep{kohl16,scha20}.

The circumbinary ring that obscures T~Tau~Sa~+~Sb is apparent in the ultraviolet (UV) image presented in Figure~\ref{fig:fig1} \citep{beck20}.  T~Tau~N is a bright point source in the UV, with an obvious circular extension of material that traces its disk.  The Hubble Space Telescope (HST) F140LP image shown in Figure~\ref{fig:fig1} reveals wispy, extended emission which is likely UV molecular hydrogen in the band pass.  To the north, molecular hydrogen is seen in disk, wind or outflow material associated with T~Tau~N.  The infrared positions of the stars are shown in red contours, and extended to the northwest and south of the position of T~Tau~Sa~+~Sb is bright UV H$_2$ flux with a 'dark lane' of no emission that traces the circumbinary ring seen in silhouette.  The decrease in line emission at the position of T~Tau~S tracing the circumbinary ring was initially hinted at in the UV spectroscopic observations of \cite{walt03}. The geometry is such that the brightest region of UV flux to the northwest is offset from the positions of the stars, which suggests that the ring extends to at least $\pm$60-80AU from the central position of Sa~+~Sb.  The ring silhouette has now been revealed in many emission species including UV molecular hydrogen (Figure~\ref{fig:fig1}), [O~I]~6300\AA~emission and 1.08$\mu$m He~I emission lines \citep{beck20}.  The mm dust associated with the circumbinary ring has been directly detected in ALMA 1.3mm maps \citep{beck20}.

“Burnham’s variable nebula” extends $\sim$10$"$ to the south of T~Tau \citep{burn90}, it has varied in morphology throughout the past century and it has an optical spectrum typical of Herbig-Haro outflows \citep{herb50}.  Interestingly, the arcs and loops of emission associated with Burnham's nebula seem to be periodic \citep{beck20}.  Analysis of the outflow arc distances from the T~Tau~System with the binary orbital timing provides striking evidence that the outflows that comprise Burnham’s nebula are likely triggered by mass accretion events caused by the close binary passage of T~Tau~Sb around Sa \citep{gust10,beck20}. 

%\cite{walt03} detected the first hint of dust obscuring the bright UV line emission at the location of T~Tau~S, which led them to postulate that orbital motion in the wide N~-~S system might obscure T~Tau~N at some point in the future.  
The motion of the N~-~S system shows that Sa~+~Sb, and the circumbinary ring, are moving in the northwestern direction \citep{kohl16,scha20}. This led both \cite{flor20} and \cite{beck20} to postulate that the northeastern side of the circumbinary ring could affect the brightness of T~Tau~N within the next $\sim$100~yr.  In this study, recent optical brightness measurements of T~Tau~N are coupled with spectral imaging data and compared to the motions of the star and disk material in this curious young triple system.  

%\latex\ \footnote{\url{http://www.latex-project.org/}} 

\section{AAVSO Light Curve and Spectral Imaging Data} \label{sec:style}

The American Association of Variable Star Observers (AAVSO) is comprised of hundreds of dedicated amateur and professional astronomical observers, and the variability database has more than 65 million observations \citep{aavs24}. Optical variability measurements of T~Tau in the AAVSO database span the 1860s to the present \citep{beck20}. The optical light curve of T~Tau from 1970 to early 2024 is shown in Figure~\ref{fig:fig2}.  This includes all measurements reported as "VIS" (no photometric filter) in the AAVSO 'photometric light curve' database for T~Tau.  A 7 day rolling smoothing window is applied to the data to even out spurious magnitudes, observer differences and the discrete measurements in 0.1 mag quanta to provide a more continuous average light curve of more than 23,000 brightness measurements.  Error bars are plotted that are determined from the standard deviation of measurements in each 7 day window used for the smoothing.  We also collected all B, R, and I photometric filter data available in the light curve database to study multi-color character of T~Tau.  In these visible band passes, the system flux is by far dominated by T~Tau~N, with less than 1\% of the light arising from any scattering associated with the T~Tau~S circumbinary ring.

Previously unpublished optical integral field spectral imaging observations of T~Tau are also presented here.  These data were acquired on 29 November 2019 with the Gemini Multi-Object Spectrograph (GMOS) Integral Field Unit (IFU) at the Gemini North Observatory on Mauna Kea (under program ID GN-2019B-Q-204; PI = Beck).  The observations used the single slit configuration with the R831 grating and RG610 filter.  This gave spectral imaging coverage from 6135\AA~to 8435\AA~with 0.378\AA/pix spectral sampling.  The Gemini IRAF processing tools for spectroscopic datasets were used to reduce the 2-Dimensional (2D) detector image to the 3-Dimensional (3D) IFU datacube with 0.$"$1 spatial sampling.  The GMOS observations were acquired with a wide scale dither pattern to map emission in the southern outflow, the but individual pointings had excellent signal-to-noise (S/N) on the emission lines and only a central pointing on T~Tau~N is presented here.  For this study, analysis of the morphology of the 6300\AA~[O~I]~emission is carried out, and compared to a nearly identical dataset acquired in 2004 and published and discussed in \cite{beck20}.  The two datasets were observed with the same instrument configuration and built to the same cube spatial sampling specification.  To compare the line emission maps consistently, the point source emission from the T~Tau~N continuum flux was fit with a 2D Gaussian at every wavelength to create a model cube of the point spread function (PSF) for the observation.  Each spectral channel was deconvolved using this PSF model to sharpen the emission and match spatial resolution in the two datasets.  The deconvolution was carried out using the iterative Bayesian-based Richardson-Lucy method \citep{rich72,lucy74}.  This robust method for sharpening an image has been shown to be both flux conserving and does not resulting in negative artifacts. The deconvolution was carried out using the python scikit image restoration tools. Figure~\ref{fig:fig4} shows the two datasets side-by-side (4a, 4b) and overlaid on each other in color (4c).

\section{Results} \label{sec:floats}

\begin{figure}[ht!]
\plotone{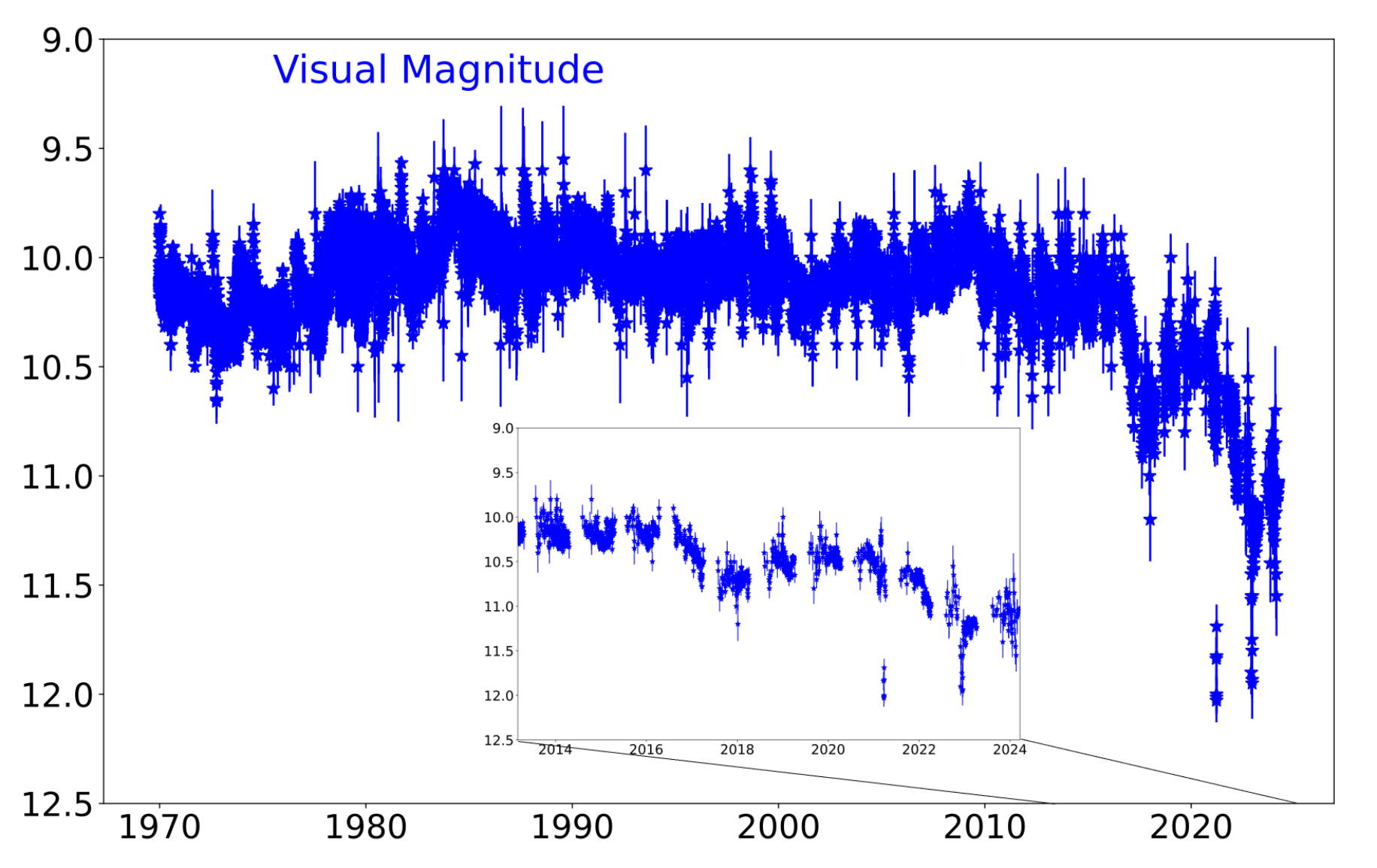}
\caption{The visual light curve of T~Tau~North from 1970 to April 2024 from over 23,000 measurements in the public AAVSO database.  The measurements have been smoothed by a 7 day rolling window (see text).  The brightness of T~Tau~N was largely stable at 10.1 $\pm$~0.3 mag from 1970 to about 2015. Over the course of the past decade, T~Tau~N has been steadily decreasing in optical brightness.   In 2016, T~Tau~N dimmed to 11.0 mag, and in 2021-2022 brief dimming events took many brightness measurements to beyond 12.0 mag in the visual.  
\label{fig:fig2}}
\end{figure}

Figure~\ref{fig:fig2} presents the visual brightness light curve of T~Tau~N from January 1970 to the present.  Prior to $\sim$2015, the brightness of T~Tau~N has been quite stable at around 10.1-10.3 magnitude in V, with a dip in the 1970's to 10.5 and a few slow dimming events by 0.1-0.2 magnitudes.   This is consistent with variability measurements from \cite{herb94}.  A few discrete decreases to 10.8-11.0 mag are seen in the light curve from 1980-2000, these are typically from spurious individual measurements of greater than 12mag that are decreased in the window smoothing.   In early 2016, T~Tau~N sharply dimmed to 11.0 mag in several measurements, and increased again to 10.5 but has not yet come back to the baseline level of 10.2-10.3 seen prior to 2015.  In 2022, several observers reported the brief dimming of T~Tau~N to beyond 12.0 mag in the visual.  The light curve in Figure~\ref{fig:fig2} shows that the optical star in the T~Tau~System has dimmed by up to 2 mag since the early 2010s, and on average is fainter by more than 1 magnitude than it was a decade ago.  Dimming of T~Tau by 2 magnitudes in the optical has not been seen in the past $\sim$100 years \citep{beck01}.

\begin{figure}[ht!]
\includegraphics[scale=0.3]{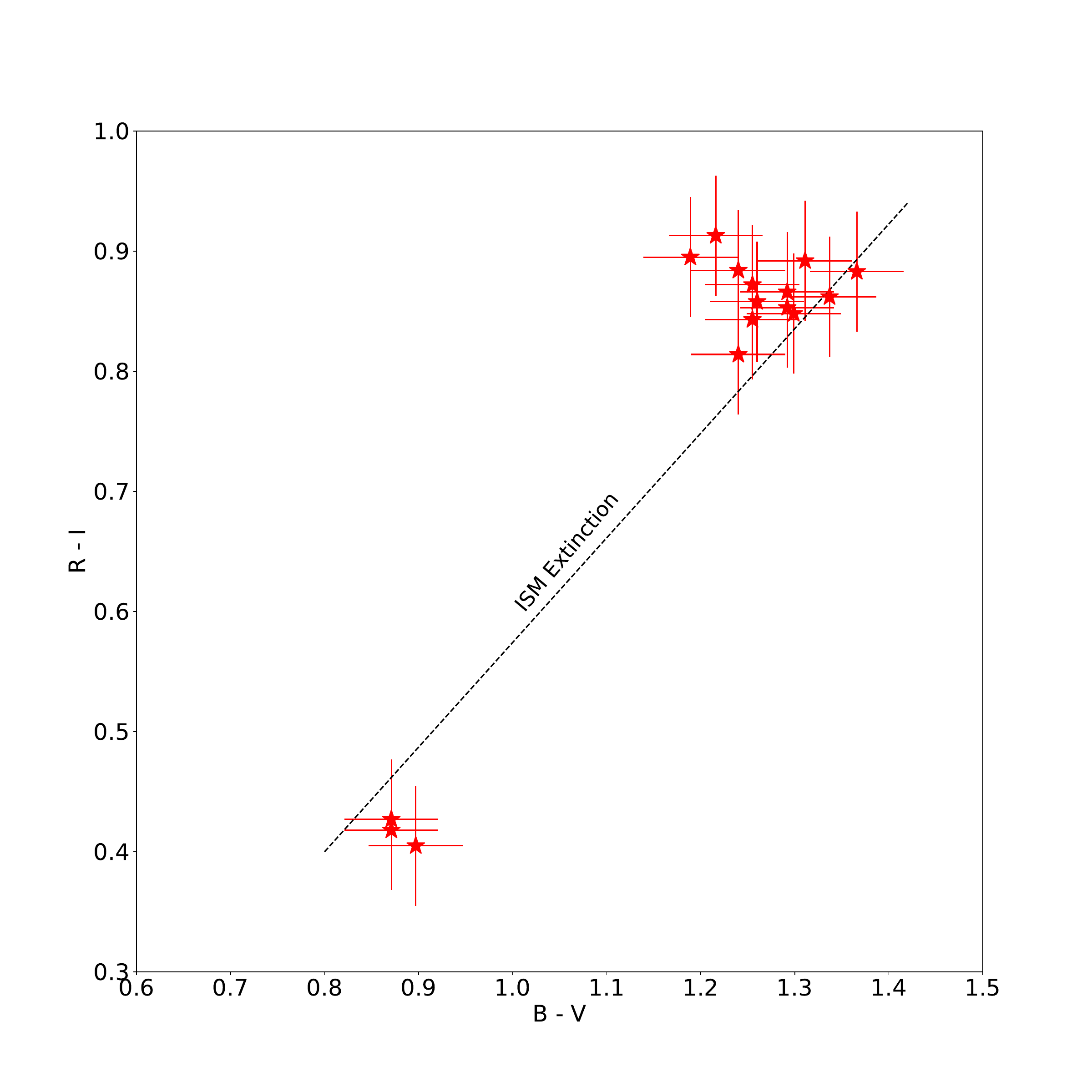}
\caption{The plot of B - V versus R - I color of T~Tau~North, from  near simultaneous (within 1 day) measures across these four bands.  Based on the available measurements from the AAVSO from 2008 to 2024. The over-plotted line shows the color changes that would result from variations in line-of-sight obscuring material that follows an ISM extinction law \citep{card89, math90}.
\label{fig:fig3}}
\end{figure}

Starting in 2008, several observers from the AAVSO reported brightness measurements of T~Tau~N in the B, V, R and I photometric pass bands.   Measurements taken within $\pm$1 day of each other are considered contemporaneous for this analysis. The multi-filter measurements allow for analysis of the color variations seen as the brightness of T~Tau~N changes.  Figure~\ref{fig:fig3} presents a color-color diagram of the B - V color plotted versus R - I.  Only three multi-filter measurements were made with T~Tau~N in a brighter state (lower left in Figure 3) in the 2008-2009 time frame, and all other measures were made when T~Tau~N was fainter and since $\sim$2018 (upper right in Figure~\ref{fig:fig3}).  Error bars of $\pm$0.05 mag are included on each point.  AAVSO observers reported uncertainties on individual filter band pass measurements that were typically at the few milli-magnitude levels (0.003 - 0.012 mag), so the included errors in Figure~\ref{fig:fig3} are very conservative.  Over plotted in Figure~\ref{fig:fig3} is a line that shows the slope of brightness variations caused by a change in obscuring material that follows an interstellar medium (ISM) extinction law \citep{math90}.  The change in brightness seen in T~Tau~N follows a "redder when faint" relation, with a color character consistent with changes in obscuration by material that follows the extinction law very closely \citep{card89, math90}.

\begin{figure}[ht!]
\plotone{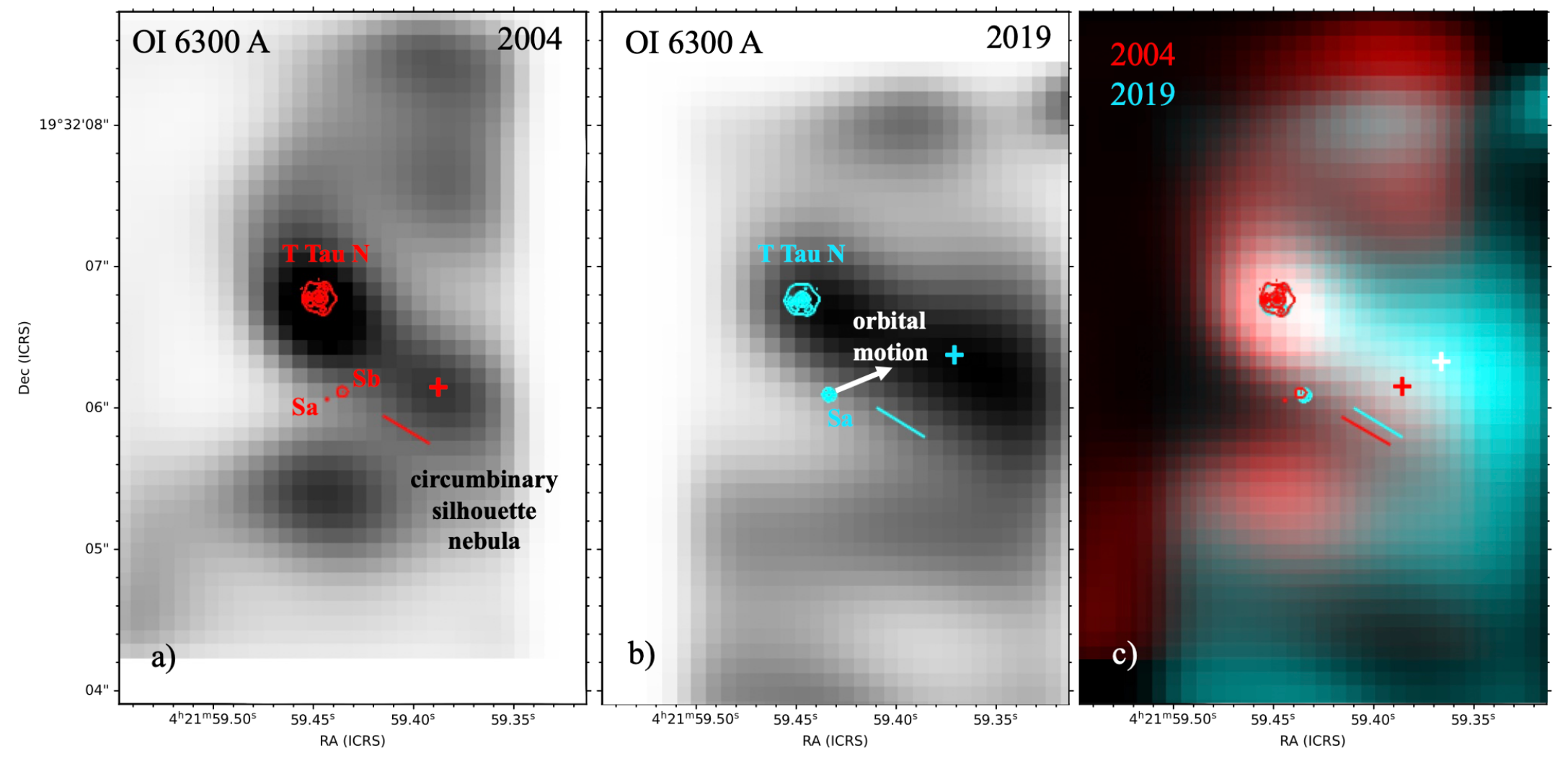}
\caption{Continuum subtracted images of the integrated 6300\AA~[O~I] line emission from the environment of the T~Tauri triple system observed in 2004 (left) and 2019 (center).  The positions of the three stars at these observing epochs from high resolution infrared measurements are shown as over-plotted contours in red and cyan, respectively \citep{beck20}.  The direction of the motion of the T~Tau~S binary in its wide orbit around T~Tau~N is shown by the arrow in the 2019 epoch image (center).  Both images are presented in color in the right panel, showing the spatial shift of the bright [O~I] emission associated with the lobes around T~Tau~S toward the north-west, in the direction of the orbital motion.
\label{fig:fig4}}
\end{figure}

The low velocity [O~I] 6300\AA~emission from young stars is known to arise from the inner disk regions, either from the disk itself or from low velocity winds \citep{rigl13}).  Figure~\ref{fig:fig4} shows the [O~I] 6300\AA~emission in the environment of T~Tau~N and S from two epochs: 2004 and fifteen years later in 2019.  Both epoch images in Figure~\ref{fig:fig4} are referenced to the 2004 position of T~Tau~N.  T~Tau~N appears to have dimmed relatively in [O~I] in 2019 compared to the lobes emission from T~Tau~S (Figure~\ref{fig:fig4}b), and also appears dimmer in [O~I] compared to its measurement in 2004 (Figure~\ref{fig:fig4}a).  The [O~I] emission from T~Tau~S is fainter than that associated with T~Tau~N in the 2004 observation, and it exhibits a dual lobe morphology that traces the circumbinary silhouette nebula seen in the UV in Figure~\ref{fig:fig1}, as expected from an origin in the inner wind.  In 2019, the northern lobe of [O~I] emission from T~Tau~S was brighter with respect to the southern lobe, whereas both were nearly equal brightness in 2004.  {\ bf The southern lobe of [O~I] emission in the 2019 epoch has dimmed and become much less distinct.}  In both epochs and in the UV image (Figure~\ref{fig:fig1}), the center of the dark lane tracing the circumbinary ring is offset from the infrared positions of the stars, which \cite{beck20} postulates to be an effect of viewing geometry.

The orbital motion of T~Tau~S with respect to T~Tau~N is progressing in a west by northwest direction, toward position angle (east of north) of $\sim$300$^{\circ}$ \citep{kohl16, scha20}.  This is seen in a directional arrow in Figure~\ref{fig:fig4}b, and is apparent by the motion of [O~I] line emission seen in the dual epoch overlay presented in Figure~\ref{fig:fig4}c.   Figure~\ref{fig:fig4} also shows the peak of the lobe emissions as asterisks, and the half-light position as a line which traces the inner edge location of the northwestern lobe in each epoch. These are also overplotted in the combined color image.  The brightness peak of the [O~I] emission has shifted further north-west in the 2019 epoch, implying a stronger, more extended wind.  Still, the leading side of the circumbinary silhouette nebula around T~Tau~S, traced by the half-light positions in Figure~\ref{fig:fig4}, has moved by $\sim$150~milliarcseconds (mas) in the direction of the orbital motion in 2019 (cyan) versus 2004 (red) in Figure~\ref{fig:fig4}c.  This suggests that the circumbinary distribution of material traced by the dark region between the two lobes of [O~I] emission is moving into our line of sight toward T~Tau~N, as completely expected by the overall wide-field orbital motion of T~Tau~S. 

%Assuming the circumbinary ring material is symmetric around the position of T~Tau~S, an extended distribution of ring material to the north-east may be causing the recent brightness dimming events seen in the optical flux since 2016, as it moves in the wide orbit.  As the T~Tau~N - S wide orbital motion continues and the circumbinary ring moves more fully into our line of sight, the dimming of T~Tau~N should increase.

%% The "ht!" tells LaTeX to put the figure "here" first, at the "top" next
%% and to override the normal way of calculating a float position

\section{Discussion}

T~Tauri was historically identified as a rapid and random variable star \citep{hind64}, which is one distinguishing characteristic that helped define the class of young sun-like stars in our galaxy.  \cite{lozi49} first collected and presented the multi-decade optical light curve of T~Tauri and showed random dimming events from the beginning of measurements in the 1860's to $\sim$1925.  \cite{beck01} and \cite{beck20} have already discussed these dimming events as changes in line of sight obscuration to T~Tau~N.   \cite{walt03} first postulated in the literature that the distribution of material associated with and attenuating T~Tau~S might at some point affect the brightness of T~Tau~N.  Such a brightness variation in T~Tau~N might start with rapid short term dimming events that will slowly increase in depth and duration.  This would occur as the denser mid-plane regions of the northeastern side of the T~Tau~S circumbinary ring move into our line of sight. 

Figure~\ref{fig:fig1} presents the T~Tau~System in the UV, 1400\AA, using the F140LP filter of the Solar Blind Channel UV array of the Advanced Camera for Surveys on the Hubble Space Telescope.  This image reveals the circumbinary ring around T~Tau~Sa+Sb in silhouette, foreground to the brighter T~Tau~N, the star + disk of which is viewed 28$^{\circ}$ from face-on \citep{beck20}.  At these UV wavelengths, light is attenuated a factor of 5 to 8 times more efficiently than in the optical \citep{card89, math90}, and the extent of the dark lane tracing the circumbinary ring is roughly 20\% larger at 0.$"$65 than revealed in the optical by the [O~I] 6300\AA~emission (0.$"$55; Figure~\ref{fig:fig4}).  Based on the relative orbital motion of the system and as revealed in Figure~\ref{fig:fig4}, the circumbinary ring around T~Tau~S is slowly moving into our line of sight to T~Tau~N and is likely the cause of the random dimming events seen in the visible light from T~Tau~N since 2016.  

The rate of orbital motion of T~Tau~Sa+Sb relative to T~Tau~N is on average about 100mas/decade (Figure~\ref{fig:fig4}, \cite{scha20}).  Hence, an annual rate of 10mas of motion as the 0.$"$65 vertical UV extent of the circumbinary ring traverses the line of sight to T~Tau~N suggests that these dimming events will continue for at least 60 to 70 years.  Moreover, the relative orbital motion is not exactly crossing the vertical extent of the circumbinary ring, but inclined at an angle (Figure~\ref{fig:fig4}).  This suggests the motion of T~Tau~N relative to the circumbinary ring will traverse a path closer to 1$"$, resulting in a longer period of dimming of up to a century or more.

T~Tau~S is observed through $\sim$20 magnitudes of visual obscuration \citep{gorh92, duch02, beck20}, believed to be predominantly from the circumbinary ring material (seen in Figures~\ref{fig:fig1} and~\ref{fig:fig4}; \cite{walt03,beck20}), though some attenuation may also be from the circumstellar disks.  The full attenuation through the whole geometric line of sight of the circumbinary ring could be as much as 40 magnitudes in the optical. This value is estimated assuming we are seeing T Tau S through the $\sim$20 magnitudes of visual extinction in the foreground region of the circumbinary ring, and we would view T Tau N through double the material in both the foreground and background portions of the ring.  The apparent projected position of T~Tau~S from T~Tau~N is $\sim$100AU (at the 143.7pc distance; \cite{bail18}), and so we will be viewing T~Tau~N through the extended material in the outer regions of the circumbinary ring.  If we view T Tau N through a longer path length in this outer region of the ring, the level of extinction might be greater than 40 magnitudes.  Ultimately, the amount of material that obscures T~Tau~N and the level to which it is attenuated depends on the density profile of this outer ring material and the path length our line of sight traverses.  Monitoring the spectra of T~Tau~N during the line of sight circumbinary ring crossing could provide an exciting opportunity to study absorption features from dust and ice in this material, which would be an analogous formation region to Kuiper belt and outer solar system bodies.

\section{Summary}
\begin{itemize}
  \item The optical star in the T~Tauri triple system, T~Tau~N, has undergone dimming events of up to 2 mag in the visual in the past 9 years, from 2015 to the present.
  \item The dimming follows a redder-when-faint character, reminiscent of changes in line of sight material that follows an ISM extinction law.
  \item Orbital motion of the southern binary, T~Tau~Sa+Sb, around T~Tau~N is causing the circumbinary ring material to move into our line of sight, resulting in the recent dimming events of the optical star.
  \item Given the extent and motion of the circumbinary ring, T~Tau~N may experience dimming events from this occultation for 60 to 70 years or more, depending on the extent of the obscuring material.
  \item As the relative wide scale orbital motion of T~Tau~Sa+Sb and N continues, denser regions of the circumbinary ring mid-plane could cross the line of sight to T~Tau~N.  This might cause up to 30-40 magnitudes of visual extinction toward the optical star, depending on the extended structure of the circumbinary ring.  Hence, T~Tau~N may temporarily disappear from the optical night sky and become an obscured, infrared-bright T~Tauri star, like the southern infrared binary.
  \item Future study of the dust and ice absorption from the line-of-sight material occulting T~Tau~N will provide a direct probe of the chemistry and structure of extended disk material, at $\sim$100 AU distance from the T~Tau~S parent stars in a region analogous to the forming Kuiper belt in our own solar system.
\end{itemize}

%% IMPORTANT! The old "\acknowledgment" command has be depreciated. It was
%% not robust enough to handle our new dual anonymous review requirements and
%% thus been replaced with the acknowledgment environment. If you try to 
%% compile with \acknowledgment you will get an error print to the screen
%% and in the compiled pdf.
%% 
%% Also note that the akcnowlodgment environment does not support long amounts of text. If you have a lot of people and institutions to acknowledge, do not use this command. Instead, create a new \section{Acknowledgments}.
\begin{acknowledgments}
I am grateful to Michal Simon and Gail Schaefer for proofreading the
draft manuscript prior to submission and providing valuable comments,
and to the anonymous referee for their careful review and assistance
in improving the manuscript.  I acknowledge with thanks the variable star observations from the AAVSO International Database contributed by observers worldwide and used in this research.  Without the extensive efforts of observers from the AAVSO, this investigation would not have been possible.  Previously unpublished data presented in this study was acquired at the Gemini Observatory under program ID GN-2019B-Q-204. Based on observations obtained at the international Gemini Observatory, a program of NSF NOIRLab, which is managed by the Association of Universities for Research in Astronomy (AURA) under a cooperative agreement with the U.S. National Science Foundation on behalf of the Gemini Observatory partnership: the U.S. National Science Foundation (United States), National Research Council (Canada), Agencia Nacional de Investigaciny Desarrollo (Chile), Ministerio de Ciencia, Tecnologia e Innovacion (Argentina), Ministrio da Cincia, Tecnologia, Inovaes e Comunicaes (Brazil), and Korea Astronomy and Space Science Institute (Republic of Korea).   Some of the data presented in this article were obtained from the Mikulski Archive for Space Telescopes (MAST) at the Space Telescope Science Institute. The specific observations analyzed can be accessed via \dataset[doi: 10.17909/0eh1-zv67]{https://doi.org/10.17909/0eh1-zv67}. 
\end{acknowledgments}

%% To help institutions obtain information on the effectiveness of their 
%% telescopes the AAS Journals has created a group of keywords for telescope 
%% facilities.
%
%% Following the acknowledgments section, use the following syntax and the
%% \facility{} or \facilities{} macros to list the keywords of facilities used 
%% in the research for the paper.  Each keyword is check against the master 
%% list during copy editing.  Individual instruments can be provided in 
%% parentheses, after the keyword, but they are not verified.

\vspace{5mm}
\facilities{AAVSO, Gemini Observatory (GMOS), HST(ACS)}

%% Similar to \facility{}, there is the optional \software command to allow 
%% authors a place to specify which programs were used during the creation of 
%% the manuscript. Authors should list each code and include either a
%% citation or url to the code inside ()s when available.

\software{astropy \citep{2013A&A...558A..33A,2018AJ....156..123A},  
          Gemini IRAF reduction \citep{2016ascl.soft08006G}, 
          scikit-image \citep{vand14}
          }

%% Appendix material should be preceded with a single \appendix command.
%% There should be a \section command for each appendix. Mark appendix
%% subsections with the same markup you use in the main body of the paper.

%% Each Appendix (indicated with \section) will be lettered A, B, C, etc.
%% The equation counter will reset when it encounters the \appendix
%% command and will number appendix equations (A1), (A2), etc. The
%% Figure and Table counter will not reset.

%% For this sample we use BibTeX plus aasjournals.bst to generate the
%% the bibliography. The sample631.bib file was populated from ADS. To
%% get the citations to show in the compiled file do the following:
%%
%% pdflatex sample631.tex
%% bibtext sample631
%% pdflatex sample631.tex
%% pdflatex sample631.tex

%\bibliography{sample631}{}

%\addbibresource{beck24bib}
\bibliography{beck24}%

\begin{thebibliography}{}
\expandafter\ifx\csname natexlab\endcsname\relax\def\natexlab#1{#1}\fi
\providecommand{\url}[1]{\href{#1}{#1}}
\providecommand{\dodoi}[1]{doi:~\href{http://doi.org/#1}{\nolinkurl{#1}}}
\providecommand{\doeprint}[1]{\href{http://ascl.net/#1}{\nolinkurl{http://ascl.net/#1}}}
\providecommand{\doarXiv}[1]{\href{https://arxiv.org/abs/#1}{\nolinkurl{https://arxiv.org/abs/#1}}}

\bibitem[{aavso.org(2024)}]{aavs24}
aavso.org. 2024, American Association of Variable Star Observers.
\newblock \url{www.aavso.org}

\bibitem[{{Astropy Collaboration} {et~al.}(2013){Astropy Collaboration},
  {Robitaille}, {Tollerud}, {Greenfield}, {Droettboom}, {Bray}, {Aldcroft},
  {Davis}, {Ginsburg}, {Price-Whelan}, {Kerzendorf}, {Conley}, {Crighton},
  {Barbary}, {Muna}, {Ferguson}, {Grollier}, {Parikh}, {Nair}, {Unther},
  {Deil}, {Woillez}, {Conseil}, {Kramer}, {Turner}, {Singer}, {Fox}, {Weaver},
  {Zabalza}, {Edwards}, {Azalee Bostroem}, {Burke}, {Casey}, {Crawford},
  {Dencheva}, {Ely}, {Jenness}, {Labrie}, {Lim}, {Pierfederici}, {Pontzen},
  {Ptak}, {Refsdal}, {Servillat}, \& {Streicher}}]{2013A&A...558A..33A}
{Astropy Collaboration}, {Robitaille}, T.~P., {Tollerud}, E.~J., {et~al.} 2013,
  \aap, 558, A33, \dodoi{10.1051/0004-6361/201322068}

\bibitem[{{Astropy Collaboration} {et~al.}(2018){Astropy Collaboration},
  {Price-Whelan}, {Sip{\H{o}}cz}, {G{\"u}nther}, {Lim}, {Crawford}, {Conseil},
  {Shupe}, {Craig}, {Dencheva}, {Ginsburg}, {VanderPlas}, {Bradley},
  {P{\'e}rez-Su{\'a}rez}, {de Val-Borro}, {Aldcroft}, {Cruz}, {Robitaille},
  {Tollerud}, {Ardelean}, {Babej}, {Bach}, {Bachetti}, {Bakanov}, {Bamford},
  {Barentsen}, {Barmby}, {Baumbach}, {Berry}, {Biscani}, {Boquien}, {Bostroem},
  {Bouma}, {Brammer}, {Bray}, {Breytenbach}, {Buddelmeijer}, {Burke},
  {Calderone}, {Cano Rodr{\'\i}guez}, {Cara}, {Cardoso}, {Cheedella}, {Copin},
  {Corrales}, {Crichton}, {D'Avella}, {Deil}, {Depagne}, {Dietrich}, {Donath},
  {Droettboom}, {Earl}, {Erben}, {Fabbro}, {Ferreira}, {Finethy}, {Fox},
  {Garrison}, {Gibbons}, {Goldstein}, {Gommers}, {Greco}, {Greenfield},
  {Groener}, {Grollier}, {Hagen}, {Hirst}, {Homeier}, {Horton}, {Hosseinzadeh},
  {Hu}, {Hunkeler}, {Ivezi{\'c}}, {Jain}, {Jenness}, {Kanarek}, {Kendrew},
  {Kern}, {Kerzendorf}, {Khvalko}, {King}, {Kirkby}, {Kulkarni}, {Kumar},
  {Lee}, {Lenz}, {Littlefair}, {Ma}, {Macleod}, {Mastropietro}, {McCully},
  {Montagnac}, {Morris}, {Mueller}, {Mumford}, {Muna}, {Murphy}, {Nelson},
  {Nguyen}, {Ninan}, {N{\"o}the}, {Ogaz}, {Oh}, {Parejko}, {Parley}, {Pascual},
  {Patil}, {Patil}, {Plunkett}, {Prochaska}, {Rastogi}, {Reddy Janga},
  {Sabater}, {Sakurikar}, {Seifert}, {Sherbert}, {Sherwood-Taylor}, {Shih},
  {Sick}, {Silbiger}, {Singanamalla}, {Singer}, {Sladen}, {Sooley},
  {Sornarajah}, {Streicher}, {Teuben}, {Thomas}, {Tremblay}, {Turner},
  {Terr{\'o}n}, {van Kerkwijk}, {de la Vega}, {Watkins}, {Weaver}, {Whitmore},
  {Woillez}, {Zabalza}, \& {Astropy Contributors}}]{2018AJ....156..123A}
{Astropy Collaboration}, {Price-Whelan}, A.~M., {Sip{\H{o}}cz}, B.~M., {et~al.}
  2018, \aj, 156, 123, \dodoi{10.3847/1538-3881/aabc4f}

\bibitem[{{Bailer-Jones} {et~al.}(2018){Bailer-Jones}, {Rybizki}, {Fouesneau},
  {Mantelet}, \& {Andrae}}]{bail18}
{Bailer-Jones}, C.~A.~L., {Rybizki}, J., {Fouesneau}, M., {Mantelet}, G., \&
  {Andrae}, R. 2018, \aj, 156, 58, \dodoi{10.3847/1538-3881/aacb21}

\bibitem[{Beck {et~al.}(2020)Beck, Schaefer, Guilloteau, Simon, Dutrey, Folco,
  \& Chapillon}]{beck20}
Beck, T.~L., Schaefer, G.~H., Guilloteau, S., {et~al.} 2020, The Astrophysical
  Journal, 902, 132, \dodoi{10.3847/1538-4357/abb5f5}

\bibitem[{{Beck} \& {Simon}(2001)}]{beck01}
{Beck}, T.~L., \& {Simon}, M. 2001, \aj, 122, 413, \dodoi{10.1086/321133}

\bibitem[{{Burnham}(1890)}]{burn90}
{Burnham}, S.~W. 1890, \mnras, 51, 94, \dodoi{10.1093/mnras/51.2.94}

\bibitem[{Cardelli {et~al.}(1989)Cardelli, Clayton, \& Mathis}]{card89}
Cardelli, J.~A., Clayton, G.~C., \& Mathis, J.~S. 1989, The Astrophysical
  Journal, 345, 245, \dodoi{10.1086/167900}

\bibitem[{{Duch{\^e}ne} {et~al.}(2002){Duch{\^e}ne}, {Ghez}, \&
  {McCabe}}]{duch02}
{Duch{\^e}ne}, G., {Ghez}, A.~M., \& {McCabe}, C. 2002, \apj, 568, 771,
  \dodoi{10.1086/338987}

\bibitem[{{Dyck} {et~al.}(1982){Dyck}, {Simon}, \& {Zuckerman}}]{dyck82}
{Dyck}, H.~M., {Simon}, T., \& {Zuckerman}, B. 1982, \apjl, 255, L103,
  \dodoi{10.1086/183778}

\bibitem[{{Flores} {et~al.}(2020){Flores}, {Reipurth}, \& {Connelley}}]{flor20}
{Flores}, C., {Reipurth}, B., \& {Connelley}, M.~S. 2020, \apj, 898, 109,
  \dodoi{10.3847/1538-4357/ab9e67}

\bibitem[{{Gemini Observatory} \& {AURA}(2016)}]{2016ascl.soft08006G}
{Gemini Observatory}, \& {AURA}. 2016, {Gemini IRAF: Data reduction software
  for the Gemini telescopes}, Astrophysics Source Code Library, record
  ascl:1608.006

\bibitem[{Gorham {et~al.}(1992)Gorham, Ghez, Haniff, Kulkarni, Matthews, \&
  Neugebauer}]{gorh92}
Gorham, P.~W., Ghez, A.~M., Haniff, C.~A., {et~al.} 1992, The Astronomical
  Journal, 103, 953, \dodoi{10.1086/116117}

\bibitem[{{Gustafsson} {et~al.}(2010){Gustafsson}, {Kristensen}, {Kasper}, \&
  {Herbst}}]{gust10}
{Gustafsson}, M., {Kristensen}, L.~E., {Kasper}, M., \& {Herbst}, T.~M. 2010,
  \aap, 517, A19, \dodoi{10.1051/0004-6361/200913828}

\bibitem[{{Herbig}(1950)}]{herb50}
{Herbig}, G.~H. 1950, \apj, 111, 11, \dodoi{10.1086/145232}

\bibitem[{{Herbst} {et~al.}(1994){Herbst}, {Herbst}, {Grossman}, \&
  {Weinstein}}]{herb94}
{Herbst}, W., {Herbst}, D.~K., {Grossman}, E.~J., \& {Weinstein}, D. 1994, \aj,
  108, 1906, \dodoi{10.1086/117204}

\bibitem[{{Hind}(1864)}]{hind64}
{Hind}, J.~R. 1864, \mnras, 24, 65

\bibitem[{{K{\"o}hler} {et~al.}(2016){K{\"o}hler}, {Kasper}, {Herbst},
  {Ratzka}, \& {Bertrang}}]{kohl16}
{K{\"o}hler}, R., {Kasper}, M., {Herbst}, T.~M., {Ratzka}, T., \& {Bertrang},
  G.~H.~M. 2016, \aap, 587, A35, \dodoi{10.1051/0004-6361/201527125}

\bibitem[{{Koresko}(2000)}]{kore00}
{Koresko}, C.~D. 2000, \apjl, 531, L147, \dodoi{10.1086/312543}

\bibitem[{Lozinskii(1949)}]{lozi49}
Lozinskii, A.~M. 1949

\bibitem[{{Lucy}(1974)}]{lucy74}
{Lucy}, L.~B. 1974, \aj, 79, 745, \dodoi{10.1086/111605}

\bibitem[{Mathis(1990)}]{math90}
Mathis, J.~S. 1990, Annual Review of Astronomy and Astrophysics, 28,
  37{\^a}70, \dodoi{10.1146/annurev.aa.28.090190.000345}

\bibitem[{{Richardson}(1972)}]{rich72}
{Richardson}, W.~H. 1972, Journal of the Optical Society of America
  (1917-1983), 62, 55

\bibitem[{Rigliaco {et~al.}(2013)Rigliaco, Pascucci, Gorti, Edwards, \&
  Hollenbach}]{rigl13}
Rigliaco, E., Pascucci, I., Gorti, U., Edwards, S., \& Hollenbach, D. 2013, The
  Astrophysical Journal, 772, 60, \dodoi{10.1088/0004-637x/772/1/60}

\bibitem[{{Schaefer} {et~al.}(2020){Schaefer}, {Beck}, {Prato}, \&
  {Simon}}]{scha20}
{Schaefer}, G.~H., {Beck}, T.~L., {Prato}, L., \& {Simon}, M. 2020, \aj, 160,
  35, \dodoi{10.3847/1538-3881/ab93be}

\bibitem[{van~der Walt~S. {et~al.}(2014)van~der Walt~S., L., J., F., D., N.,
  E., T., \& the scikit-image contributors}]{vand14}
van~der Walt~S., L., S.~J., J., N.-I., {et~al.} 2014, scikit-image: image
  processing in Python.

\bibitem[{{Walter} {et~al.}(2003){Walter}, {Herczeg}, {Brown}, {Ardila},
  {Gahm}, {Johns-Krull}, {Lissauer}, {Simon}, \& {Valenti}}]{walt03}
{Walter}, F.~M., {Herczeg}, G., {Brown}, A., {et~al.} 2003, \aj, 126, 3076,
  \dodoi{10.1086/379557}

\end{thebibliography}
\bibliographystyle{aasjournal}

%\printbibliography
%van der Walt S, Schönberger JL, Nunez-Iglesias J, Boulogne F, Warner JD, Yager N, Gouillart E, Yu T, the scikit-image contributors. 2014. scikit-image: image processing in Python. PeerJ 2:e453 https://doi.org/10.7717/peerj.453

%\bibliographystyle{aasjournal}

%% This command is needed to show the entire author+affiliation list when
%% the collaboration and author truncation commands are used.  It has to
%% go at the end of the manuscript.
%\allauthors

%% Include this line if you are using the \added, \replaced, \deleted
%% commands to see a summary list of all changes at the end of the article.
%\listofchanges

\end{document}